\documentclass[authoryear,preprint,review,12pt]{elsarticle}

\usepackage{graphics}
\usepackage{graphicx}
\usepackage{epsfig}
\usepackage{epstopdf}
\usepackage{amssymb}
\usepackage{amsmath}
\usepackage{amsthm}
\usepackage{color}
\usepackage{lineno}
\usepackage{float}
\usepackage[table]{xcolor}
\usepackage{multirow}

\begin{document}



\centerline{\normalsize \textbf{Combination of institutional incentives for cooperative governance of risky commons}}
\vbox{}
\centerline{Weiwei Sun$^{1,\dag}$, Linjie Liu$^{1,\dag}$, Xiaojie Chen$^{1,5,*}$, Attila Szolnoki$^2$, and V\'{\i}tor V. Vasconcelos$^{3,4}$}
\vbox{}

$^1$School of Mathematical Sciences, University of
Electronic Science and Technology of China, Chengdu 611731, China

$^2$Institute of Technical Physics and Materials Science, Centre for Energy Research, P.O. Box 49, Budapest H-1525, Hungary

$^3$Informatics Institute, University of Amsterdam, Science Park 904, 1098XH Amsterdam, The Netherlands

$^4$Institute for Advanced Study, University of Amsterdam, 1012 GC, Amsterdam, The Netherlands

$^5$ Lead Contact

$^{\dag}$ These authors contributed equally to this study.

$^*$ Correspondence: xiaojiechen@uestc.edu.cn (X.C.)

\newpage
\section*{SUMMARY}
Finding appropriate incentives to enforce collaborative efforts for governing the commons in risky situations is a long-lasting challenge. Previous works have demonstrated that both punishing free-riders and rewarding cooperators could be potential tools to reach this goal. Despite weak theoretical foundations, policy-makers frequently impose a punishment-reward combination. Here, we consider the emergence of positive and negative incentives and analyze their simultaneous impact on sustaining risky commons. Importantly, we consider institutions with fixed and flexible incentives. We find that a local sanctioning scheme with pure reward is the optimal incentive strategy. It can drive the entire population towards a highly cooperative state in a broad range of parameters, independently of the type of institutions. We show that our finding is also valid for flexible incentives in the global sanctioning scheme, although the local arrangement works more effectively.


\section*{INTRODUCTION}

\noindent Our society faces many urgent challenges that can be encapsulated as ``the problems of the commons", such as climate change, desertification of land, environmental pollution, and the list can be easily extended by other examples~\citep{ostrom_90}. All these pressing problems require us to give up our short-term personal interests for the benefit of long-term collective success ~\citep{rand_tcs13,tavoni_ncc13,lenton_ncc14,Shirado_20}.
However, self-interested individuals always prefer their personal interests which can easily lead to a collective failure, hence the ``tragedy of the commons" seems to be inevitable~\citep{hardin_science68}.

\hspace*{\fill}

\noindent As a paradigm, the public goods game has been commonly used to study such dilemmas characterized by the frustration between personal and collective interests~\citep{fehr_nature02,hauert_jtb02,szolnoki_pre11,sasaki_pnas12,chen_jrsi15,han_jrsif15,wang_cns19,ginsberg_games19,Domingos_is20,Santos_is21}.  However, recent research emphasized that the traditional public goods game does not consider the risk of group failure,
hence the so-called collective-risk social dilemma game is more competent tool to grab the essence of the conflict in several cases~\citep{milinski_pnas08}. An excellent example of this feature is the mitigation effort to prevent climate change~\citep{schroeder_ncc12,barrett_pnas12,marotzke_ncc20}, which is one of the greatest public goods dilemmas that we humans face in the history~\citep{inman_ncc09}.

\hspace*{\fill}

\noindent Recent experimental and theoretical studies concerning collective-risk social dilemmas have revealed that the risk of collective failure plays an important role in promoting the evolution of cooperation~\citep{milinski_pnas08,wang_pre09,wang_pre10,santos_pnas11,milinski_cc11,jacquet_ncc13,wu_plos13,zhang_sr13}. More precisely, a high risk can significantly enhance the willingness to cooperate, but cooperation collapses when the risk of failure is low~\citep{santos_mmmas12,chen_pre12,chen_pre14,pacheco_plr14,vasconcelos_pnas14,vasconcelos_mmmas15,diaz_ncc17}. Introducing costly punishment into the collective-risk social dilemma game can reverse the disadvantage situation under low risk, even when punishment is voluntary and costly~\citep{vasconcelos_ncc13}. Furthermore, local schemes of sanctioning can do better than global schemes of institutions in improving the general group achievement, even when the risk of collective failure is low \citep{vasconcelos_ncc13}.

\hspace*{\fill}

\noindent Beside punishment~\citep{sigmund_nature10}, other incentive strategies, like rewarding cooperators or the simultaneous presence of independent incentives, were also considered by previous works~\citep{hilbe_prsb10,szolnoki_prx13}. An intriguing question is whether there exists the optimal relation of negative and positive incentives that can be maintained when voluntary and costly. In this paper, we thereby consider a hybrid enforcement strategy which imposes both kinds of incentives on participants simultaneously in the collective-risk social dilemma game. Accordingly, the positive and negative incentives are simultaneously presented within an incentive strategy, called executor strategy, and executors both punish defectors and simultaneously reward cooperators with a certain weight.

\hspace*{\fill}

\noindent Our principal goal is to investigate how the proposed executor strategy influences the group achievement for solving the collective-risk social dilemma and to further find the optimal incentive strategy in the cooperative governance of risky commons when institutions are working at a local or global level. Furthermore, we respectively consider two cases of institutions: when the imposed incentive is fixed, the incentive amount is independent of the actual number of executor players; while when the imposed incentive is flexible, the incentive amount depends on the actual number of executor players
~\citep{sigmund_nature10,sasaki_jtb11,vasconcelos_ncc13,szolnoki_prsb15}. We stress that in our framework whether the institutions can be formed depends on the number of executor players in game interactions, and hence this setting is different from pervious works~\citep{chen_jrsi15,gois_sr19} which supposed that the considered centralized institutions are established in advance and can always work stably. In addition, the incentive budget on cooperators and defectors is not constrained in this work, whereas it is fixed in previous works~\citep{chen_jrsi15,gois_sr19}. Interestingly, we show that independently of the types of institutions, pure reward strategy can always do better than other combinations of incentives in improving the level of group achievement in local sanctioning scheme. Besides, we find that this observation remains valid for flexible incentives in global sanctioning scheme. Furthermore, we reveal that regardless of the value of risk, local scheme can promote group success more effectively than global scheme, no matter whether the imposed incentives are fixed or flexible.

\section*{RESULTS}

\noindent \textbf{Fixed incentives under local scheme}

\noindent We investigate the stationary distribution and the gradient of selection to study the evolutionary dynamics of cooperators ($C$), defectors ($D$), and executors ($E$) in finite well-mixed populations \citep{vasconcelos_ncc13}. The stationary distribution standing for the prevalence in time of each configuration of the entire population is mapped onto the triangular simplex, in which each dot represents a configuration. The so-called gradient of selection provides the most likely direction of evolution from a given configuration.

\begin{figure*}[htb]
\begin{center}
\includegraphics[width=12cm]{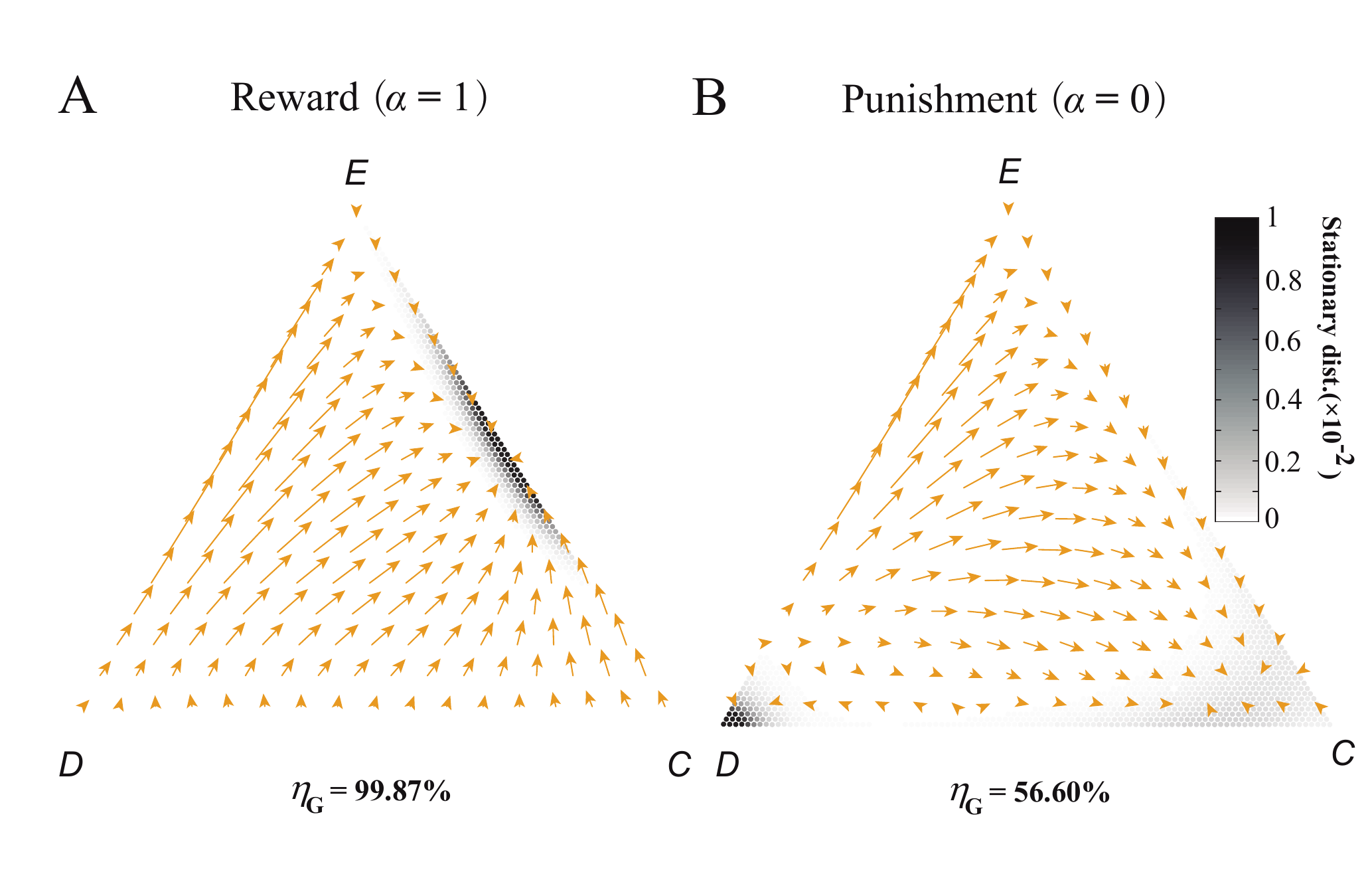}
\caption{\textbf{Evolutionary dynamics of cooperators, defectors, and executors under local scheme with fixed incentives.} Here, the risk level is \emph{r} = 0.2. In panel A, in the extreme case of $\alpha = 1$, $E$ becomes equivalent to a pure rewarding strategy, while the other extreme case of $\alpha = 0$, where $E$ is a pure punisher, can be seen in panel B. The darker dots in the simplex represent the regions where the population spends more time. Orange arrows represent the most likely direction of evolution when the population leaves the current configuration, obtained by computing the gradient of selection. We set the collective target $M$ to $75\%$ of the group size. Local scheme is established when the number of executors exceeds 25$\%$ of the group. Parameter values are \emph{Z} = 100, \emph{N} = 4, \emph{c} = 0.1, \emph{b} = 1, $\mu =1/Z$, $\pi_{t}=0.03$, and $\pi_{e}=0.3$.}\label{fig1}
\end{center}
\end{figure*}

\hspace*{\fill}

\noindent In Figure~\ref{fig1}, we show the stationary distribution and the gradient of selection when pure reward strategy (i.e., $\alpha=1$) and pure punishment strategy (i.e., $\alpha=0$) are considered respectively. It suggests that the population will spend a significant time near the \emph{CE}-edge of the triangle simplex, as shown in Figure~\ref{fig1}A. Furthermore, most of the arrows in the simplex flow to the intermediate region of $CE$-edge, indicating that the entire population evolves towards cooperation dominant states. In Figure~\ref{fig1}B we show that the population will spend most of the time near the configurations in which a lot of individuals are defectors~\citep{vasconcelos_ncc13}.
The comparison of these panels suggests that pure reward strategy can lead to better outcomes than pure punishment strategy since the population spends most of the time away from the $D$ vertex in the usage of pure reward strategy. In order to qualify this, we further compute the average group achievement $\eta_{G}$ values, which denote the average fraction of groups that succeed in achieving collective targets to measure the level of cooperation. We find that when pure reward strategy is considered, the average group achievement value is $99.87\%$. This value is much higher than the average group achievement $56.60\%$ induced by pure punishment strategy.

\begin{figure*}
\begin{center}
\includegraphics[width=13cm]{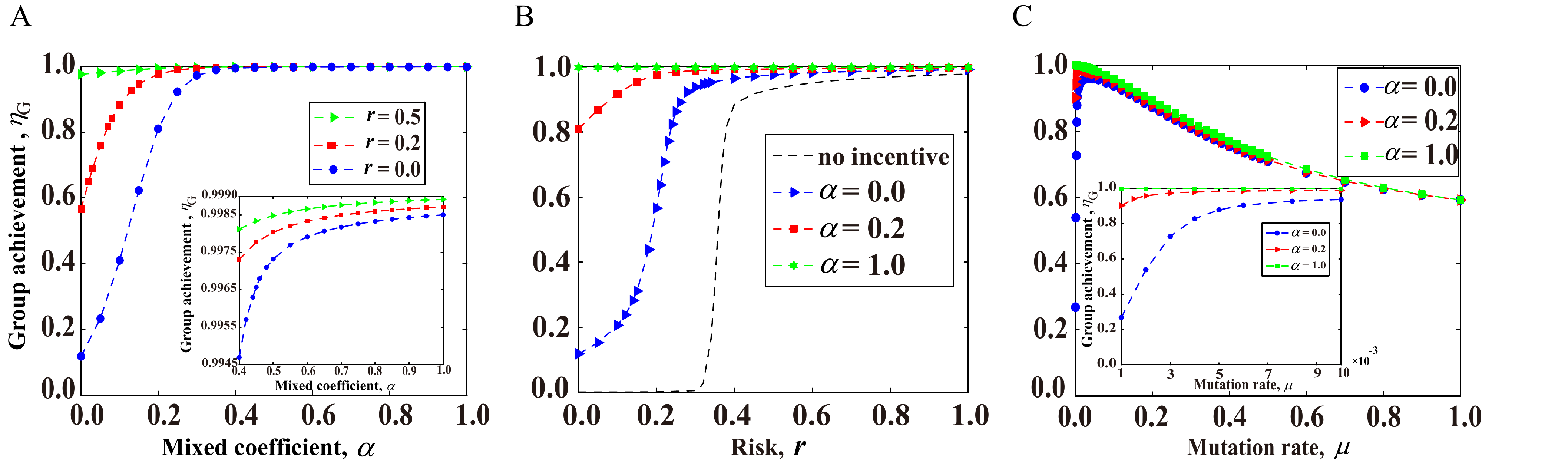}
\caption{\textbf{The average group achievement $\eta_{G}$ under local scheme with fixed incentives.} Panel A shows $\eta_{G}$ as a function of $\alpha$ for three different values of \emph{r}. The inset of panel A shows $\eta_{G}$ as a function of $\alpha$, where the value of $\alpha$ ranges from 0.4 to 1.0. Panel B shows $\eta_{G}$ as a function of \emph{r} in the cases without incentive strategy $E$ and with three different weight factor values in its presence. Panel C shows $\eta_{G}$  as a function of $\mu$ for three different values of $\alpha$. The inset of panel~\emph{c} shows $\eta_{G}$ as a function of $\mu$, where the value of $\mu$ ranges from 0.001 to 0.01. Parameter values are \emph{Z} = 100, \emph{N} = 4, \emph{c} = 0.1, \emph{b} = 1, $\mu = 1/Z$, $\pi_{t} = 0.03$, and $\pi_{e} = 0.3$ in panels~A and B; \emph{Z} = 100, \emph{N} = 4, \emph{c} = 0.1, \emph{b} = 1, \emph{r} = 0.3, $\pi_{t}=0.03$, and $\pi_{e}=0.3$ in panel~C.}\label{fig2}
\end{center}
\end{figure*}

\hspace*{\fill}

\noindent In order to further explore how the hybrid incentive strategy influences the group achievement and to find the optimal weight of incentives, we show the average group achievement $\eta_{G}$ as a function of $\alpha$ for different values of risk $r$ in Figure~\ref{fig2}A. We find that the group achievement first grows fast and then increases slowly when increasing the value of $\alpha$. Thus, we can conclude that pure reward strategy can always do better in improving the level of group achievement than other combinations of incentives. We note that here $c/b=0.1$ was used to obtain Figure~\ref{fig2}A, but our findings remain valid for broad range of $c/b$ ratio.

\hspace*{\fill}

\noindent We further present $\eta_{G}$ as a function of risk \emph{r} for different values of $\alpha$, as shown in Figure~\ref{fig2}B. For the sake of comparison, we also show the results when there is no incentive strategy (see black dash line). We find that the level of group achievement increases as we increase \emph{r}. Secondly, the introduction of incentive strategy provides better outcome compared to the traditional two-strategy case. Furthermore, when strategy $E$ is introduced, it is better to adjust the weight factor towards the pure rewarding case which can improve the result further especially in the small risk region.

\hspace*{\fill}

\noindent In addition to the risk of collective failure mentioned above, the mutation rate $\mu$ also plays an important role in the evolution of cooperation \citep{santos_pnas11,vasconcelos_ncc13}. However, it is unclear whether pure reward is still the optimal incentive strategy when individual mutation is considered as it introduces exogenous behavioral changes. To answer this question, in Figure~\ref{fig2}C we show $\eta_{G}$ as a function of $\mu$ for three different values of $\alpha$. We find that the group achievement first increases until reaching the maximum, and then it decreases with the increasing value of $\mu$ for different $\alpha$ values. We find that when the mutation value is high, the group achievement level reaches a certain value regardless of the value of $\alpha$, which corresponds to random play, but it is always the highest for $\alpha=1$ and this difference is the most conspicuous for small mutation rates (see the inset of Figure~\ref{fig2}C). Thus, we can conclude that our main finding is robust against the introduction of mutation.

\begin{figure*}
\begin{center}
\includegraphics[width=12cm]{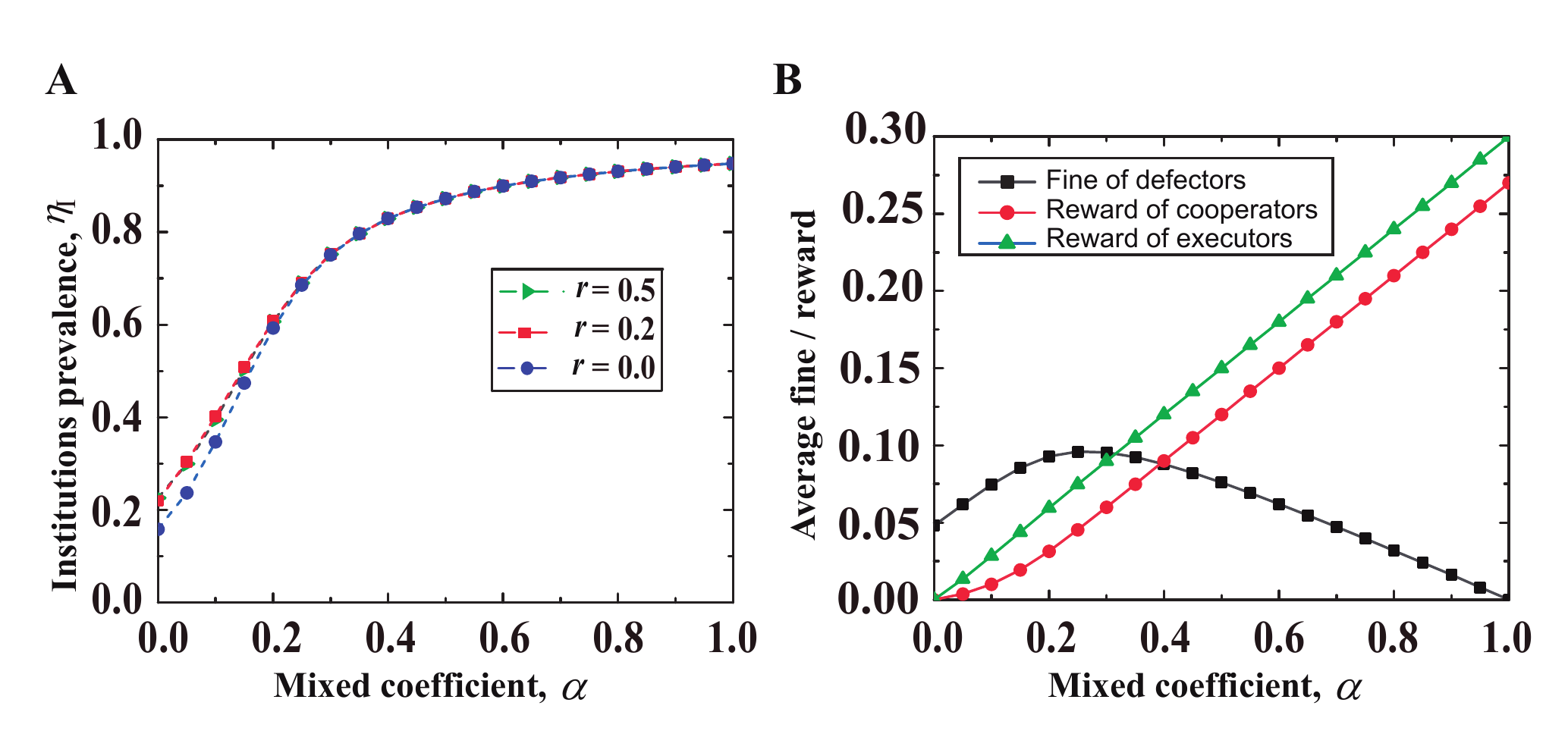}
\caption{\textbf{The institution prevalence $\eta_I$ and the average reward and fine under local scheme with fixed incentives.} Panel~A shows $\eta_I$ as a function of $\alpha$ for three different values of \emph{r}. Panel~B shows the average fine on defectors and the average reward on cooperators and executors as a function of $\alpha$. Parameter values are \emph{Z} = 100, \emph{N} = 4, \emph{c} = 0.1, \emph{b} = 1, $\mu = 1/Z$, $\pi_{t} = 0.03$, and $\pi_{e} = 0.3$ in panel~A; \emph{Z} = 100, \emph{N} = 4, \emph{c} = 0.1, \emph{b} = 1, \emph{r} = 0.2, $\mu = 1/Z$, $\pi_{t}=0.03$, and $\pi_{e}=0.3$ in panel~B.}\label{fig3}
\end{center}
\end{figure*}

\hspace*{\fill}

\noindent We further find that the success of pure rewarding strategy for fixed incentives is induced by the efficiency of the institution. As shown in Figure~\ref{fig3}A, the institution prevalence increases as we increase $\alpha$ for different risk values of $r$. We further compute the average fine on defectors and the average reward on cooperators and executors as presented in Figure~\ref{fig3}B, and observe that as the parameter $\alpha$ increases, the average reward amount on cooperators and executors monotonically increases from zero. On the other hand, the average fine amount on defectors first increases until reaching the maximum, and then decreases with increasing the value of $\alpha$. We can find that for $\alpha=1$, executors and cooperators have the greatest evolutionary advantage over defectors, which is most conductive to the institution prevalence and the group achievement.

\hspace*{\fill}

\noindent \textbf{Flexible incentives under local scheme}

\begin{figure*}
\begin{center}
\includegraphics[width=12cm]{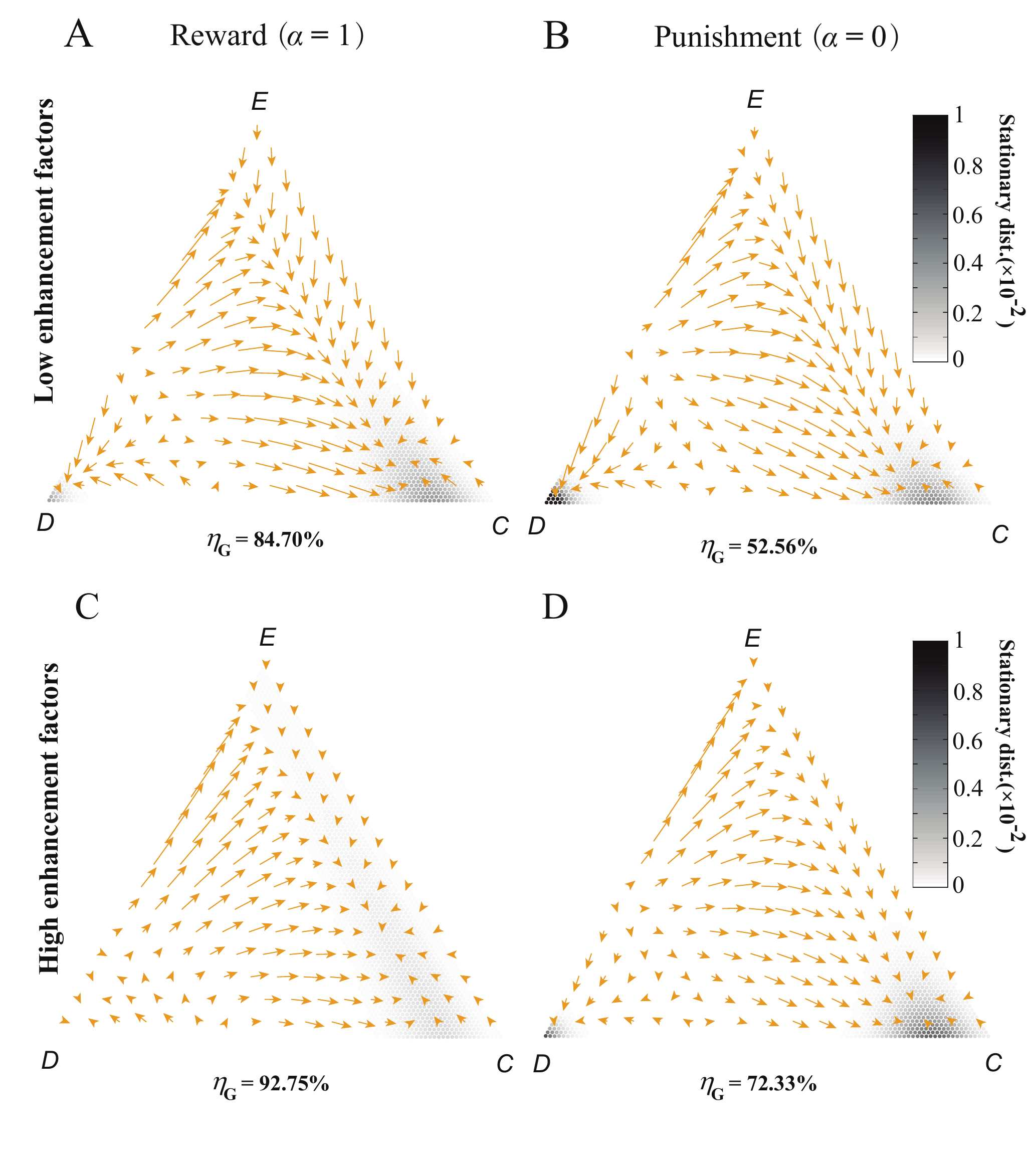}
\caption{\textbf{Evolutionary dynamics of cooperator, defector, and executor strategy under local scheme with flexible incentives.} The risk level is \emph{r} = 0.35. Panels~A and C show the case of pure reward ($\alpha=1$). Panels~B and D show the case of pure punishment ($\alpha=0$). Parameter values are \emph{Z} = 100, \emph{N} = 4, \emph{c} = 0.1, \emph{b} = 1, $\mu = 1/Z$, $\pi_{t}=0.03$, and $\delta= 1.4$ in panels A and B; \emph{Z} = 100, \emph{N} = 4, \emph{c} = 0.1, \emph{b} = 1, $\mu = 1/Z$, $\pi_{t}=0.03$, and $\delta= 3$ in panels~C and D.}\label{fig4}
\end{center}
\end{figure*}

\noindent Beside fixed incentives, we can also apply flexible values imposed by the executor strategy where the actual level depends on the number of $E$ players. Furthermore, the success of their collective efforts is acknowledged via an enhancement factor $\delta$~\citep{sasaki_jtb11,sasaki_pnas12}. Hence, an enhanced fine and reward are distributed among defecting and cooperating players respectively. In Figure~\ref{fig4}, we first show the stationary distribution and the gradient of selection for low and high enhancement factor values
in the extreme cases of $\alpha=1$ (pure reward) and $\alpha=0$ (pure punishment). As shown in Figure~\ref{fig4}A, the entire population will spend most time near configurations in which most individuals are cooperators when $\delta$ is low. Besides, we observe that most arrows flow to the vertex \emph{C} and a few arrows point to vertex \emph{D}. In Figure~\ref{fig4}B, the flow of arrows is basically consistent with that observed in Figure~\ref{fig4}A. However, the background shadow area near vertex \emph{C} in Figure~\ref{fig4}A is significantly larger than the corresponding region in Figure~\ref{fig4}B, which suggests that pure reward strategy has an evolutionary advantage over pure punishment strategy in improving the level of group achievement. For high enhancement factor of the institutional fund, the application of pure reward makes the system more cooperative and the entire population spend most of the time in the vicinity of the \emph{CE}-edge, as shown in Figure~\ref{fig4}C.
Besides, most of the arrows in the simplex flow to the regions near the $C$-corner. The corresponding simplex of pure punishment shown in Figure~\ref{fig4}D illustrates very clearly that the dark area is close to vertex $D$, hence we can conclude that the usage of pure reward strategy is more beneficial than the application of pure punishment strategy.

\hspace*{\fill}

\begin{figure*}
\begin{center}
\includegraphics[width=12cm]{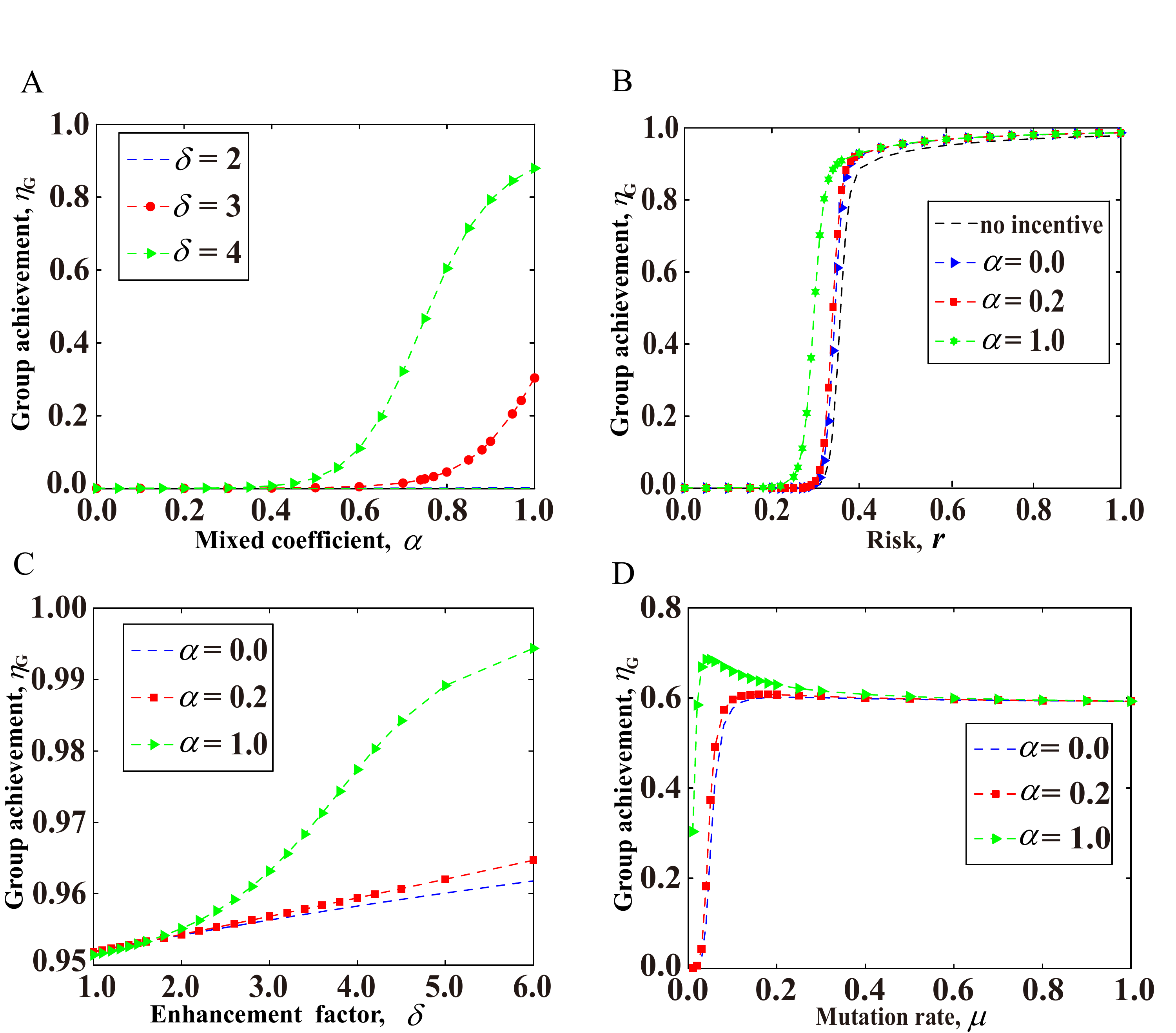}
\caption{\textbf{The average group achievement $\eta_{G}$ under local scheme with flexible incentives.} Panel A shows $\eta_{G}$ as a function of $\alpha$ for three different values of $\delta$. Panel B shows $\eta_{G}$ as a function of $r$ in the cases without incentive strategy E and with three different values of $\alpha$. Panel C shows $\eta_{G}$ as a function of $\delta$ for three different values of $\alpha$. Panel D shows $\eta_{G}$ as a function of $\mu$ for three different values of $\alpha$. Parameter values are \emph{Z} = 100, \emph{N} = 4, \emph{c} = 0.1, \emph{b} = 1, $\mu=1/Z$, $\pi_{t}=0.03$, and $r$ = 0.2 in panel A; \emph{Z} = 100, \emph{N} = 4, \emph{c} = 0.1, \emph{b} = 1, $\mu=1/Z$, $\pi_{t}=0.03$, and $\delta=2$ in panel B; \emph{Z} = 100, \emph{N} = 4, \emph{c} = 0.1, \emph{b} = 1, $\mu=1/Z$, $\pi_{t}=0.03$, and \emph{r} = 0.5 in panel~C; \emph{Z} = 100, \emph{N} = 4, \emph{c} = 0.1, \emph{b} = 1, $\mu = 1/Z$, $\pi_{t}=0.03$, \emph{r} = 0.2, and $\delta=3$ in panel~D.}\label{fig5}
\end{center}
\end{figure*}

\noindent We also compute and add the average group achievement $\eta_G$ values to the simplexes, which also illustrate clearly that pure reward performs better and this impact can be increased for higher value of enhancement factor of the institution pool. In the following, we present how the group achievement $\eta_{G}$ varies with the value of $\alpha$ for different values of $\delta$ in Figure~\ref{fig5}A. This panel demonstrates clearly that the optimal weight of incentives is $\alpha=1$, corresponding to the pure reward case. In particular, we find that for low $\delta$ (e.g., $\delta=2$), the level of group achievement approaches zero for any value of $\alpha$. However, for a larger $\delta$ value, the group achievement level grows as $\alpha$ increases and can be significantly improved when the value of $\alpha$ is larger than an intermediate value.

\hspace*{\fill}

\noindent We further present $\eta_{G}$ as a function of $r$ for different values of $\alpha$, as shown in Figure~\ref{fig5}B. We also show the results in the case when additional incentive strategy is absent (see black dash line). In general, the group achievement level increases by enhancing the risk level $r$. Besides, we observe that the introduction of the incentive strategy may elevate the value of $\eta_{G}$, especially in the pure reward extreme case, but the improvement is not as shocking as for the fixed incentives version, shown in Figure~\ref{fig2}B. In Figure~\ref{fig5}C, we show how the group achievement $\eta_{G}$ varies with increasing the enhancement factor $\delta$ for different values of $\alpha$. In general, the group achievement grows as we increase $\delta$, but the most remarkable improvement can be seen in the pure reward limit at $\alpha=1$. Finally we also explore the possible impact of mutation rate $\mu$ on the group achievement for three different values of $\alpha$. As Figure~\ref{fig5}D shows, the introduction of a small mutation rate can promote the group achievement significantly. However, when the $\mu$ becomes high, the group achievement level reaches a certain value regardless of the value of $\alpha$. But as previously, the best improvement can obtained again for pure reward strategy.

\hspace*{\fill}

\begin{figure}
\begin{center}
\includegraphics[width=12cm]{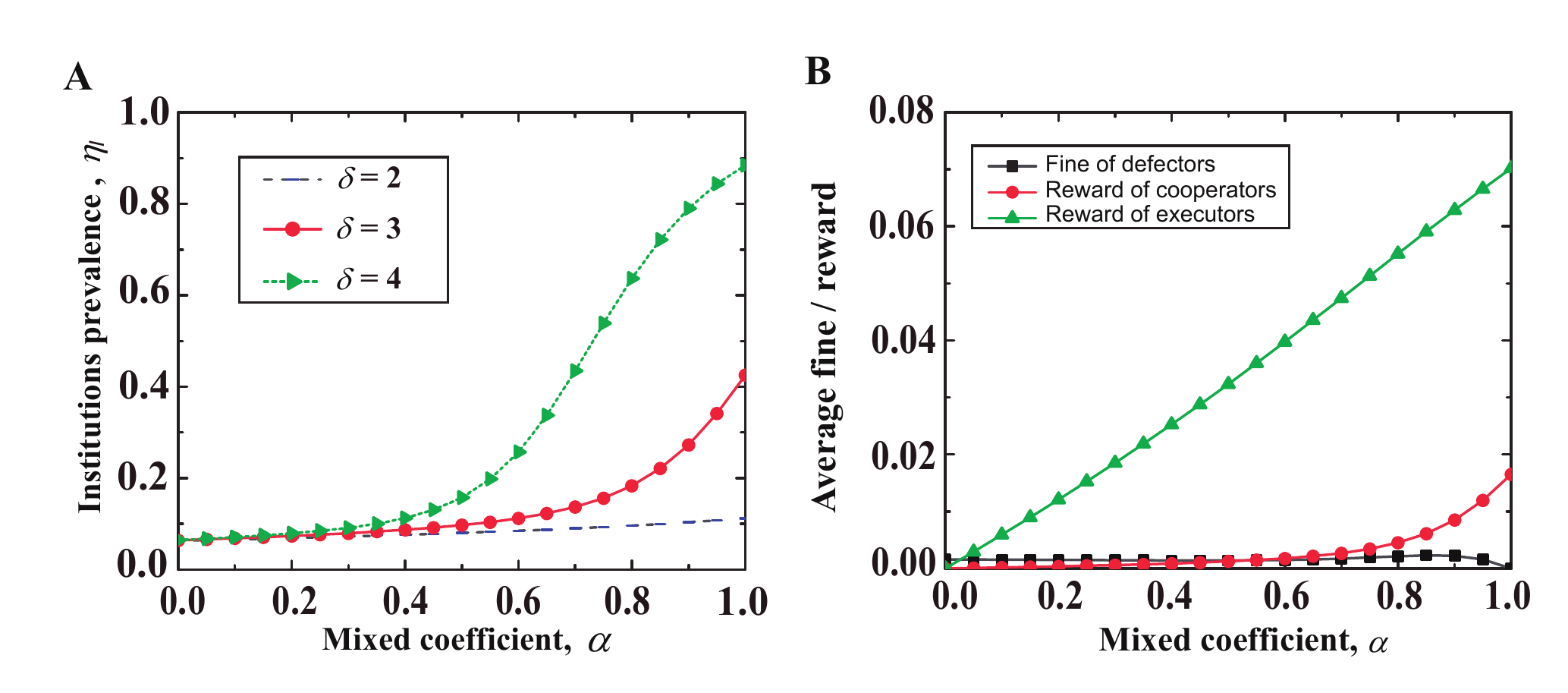}
\caption{\textbf{The institution prevalence $\eta_I$ and the average reward and fine under local scheme with flexible incentives.} Parameter values are \emph{Z} = 100, \emph{N} = 4, \emph{c} = 0.1, \emph{b} = 1, $\mu=1/Z$, $\pi_{t}=0.03$, and $r$ = 0.2 in panel~A; \emph{Z} = 100, \emph{N} = 4, \emph{c} = 0.1, \emph{b} = 1, $\mu=1/Z$, $\delta=3$, $\pi_{t}=0.03$, and $r$ = 0.2 in panel~B.}\label{fig6}
\end{center}
\end{figure}

\noindent In order to explore why pure reward strategy is more beneficial to the group achievement than other incentive strategies, we also present the institution prevalence $\eta_I$ in dependence of the weight factor $\alpha$, as shown in Figure~\ref{fig6}A. We find that for flexible incentives, the institution prevalence increases as the weight factor increases for different values of $\delta$, which means that the $\eta_I$ value reaches the maximum for $\alpha=1$. This indicates that the efficiency and prevalence of such institution can be best enhanced by the pure reward strategy. In addition, we see that the average reward amount on cooperators and executors increases from zero as we increase the $\alpha$ value (Figure~\ref{fig6}B). On the contrary, the average fine on defectors slowly decreases and reaches to zero when $\alpha$ is one (Figure~\ref{fig6}B). We then find that the evolutionary advantage of cooperators and executors over defectors can be best promoted by the pure reward strategy since their incentive difference can reach the maximum for $\alpha=1$, which can facilitate the success of group achievement.

\section*{DISCUSSION}

\noindent In this study, we have introduced a combined incentive strategy to address the collective-risk social dilemma game, where the corresponding players punish free-riders and reward cooperators simultaneously. The fundamental question is how to share available resources to execute these tasks. We have found that pure reward strategy can lead to the highest level of group achievement, no matter whether we apply fixed or flexible incentives for governing the commons. It was demonstrated that when the risk value of collective failure, the mutation rate, or the enhancement factor for institutional sanctioning is changed to some extent, our conclusion about the superiority of pure reward remains intact. Furthermore, we stress that this finding is not only valid under local scheme of sanctioning, but also valid for flexible incentives under global level of institutions (see Supplemental Information where the latter option is discussed in details). The comparison of the mentioned cases also reveals that the application of local incentives works more efficiently.

\hspace*{\fill}

\noindent Previous works have demonstrated the advantage of polycentric governance approach in solving the collective-risk social dilemma and found that local scheme of sanctioning based on pool punishment is more effective than that associated with a single, global one for providing better conditions both for cooperation to thrive and for ensuring the maintenance of such sanctioning ~\citep{vasconcelos_ncc13,green_ncc14,cole_ncc15}. In our work, we also confirm this conclusion, and we further find that it is valid not only for fixed, but also for the case when incentives are flexible. Thus, our work has greatly extended the validity of previous conclusions.

\hspace*{\fill}

\noindent In our work, the incentives on cooperators and defectors are imposed by executors, a third combined strategy introduced into the collective-risk social dilemma. This consideration of the enforcement strategy as a voluntary enforcement involves a more complex three-strategy system. Indeed, this is a kind of a bottom-up self-regulation for influencing the evolution of cooperation \citep{sugiarto_jpc17}. We stress that this approach is different from the top-down-like incentive mechanism used in previous works \citep{chen_jrsi15,gois_sr19}, in which cooperators can be rewarded and defectors can be punished directly by external centralized institutions. These works have concluded that the adaptive hybridization of incentive can best promote the evolution of cooperation both in the traditional public goods game \citep{chen_jrsi15} and in the collective-risk social dilemma game \citep{gois_sr19}. When we consider the exogenous stability of these institutions, we find that pure reward strategy can better promote the evolution of cooperation than hybrid incentive strategies in the collective-risk social dilemma. This is because in our framework the incentive budget on individuals is not constrained, and the evolutionary advantage of cooperators and executors over defectors can be significantly promoted due to the maximal incentive difference induced by the pure reward strategy. Hence, the efficiency of institution can be best enhanced by the pure reward strategy, and from an evolutionary perspective we show how an incentive strategy can be maintained as an endogenously chosen institution for cooperative governance of risky commons. Thus, our work can enrich the knowledge of the emergence of institutions for governing the commons \citep{sigmund_nature10,vasconcelos_ncc13,dannenberg_ee20}.

\hspace*{\fill}

\noindent Many researches have demonstrated that reward or punishment is a viable approach in promoting the evolution of public cooperation \citep{szolnoki_pre11,sasaki_jtb11,sasaki_pnas12}. However, identifying the best way to distribute incentives remains an open question, given the difficulty in assessing the advantages and disadvantages of each possibility in different scenarios \citep{hilbe_prsb10,szolnoki_prx13}. The majority of previous researches addressing the ``stick vs. carrot" management strategy concluded that punishment is more effective than reward in sustaining common cooperation in the conventional public goods game~\citep{sigmund_tee07,sasaki_pnas12}. However, the conventional public goods game fails to capture the significant feature of many real social scenarios, such as climate change \citep{green_ncc14}. In this work, instead of the conventional public goods game we apply the collective-risk social dilemma game which has been proved as a framework to investigate the inherent problems of collective cooperation regarding avoiding dangerous climate change and other problems of this type \citep{santos_pnas11}, and we show that the optimal sanctioning policy is pure reward, no matter whether fixed or flexible incentives are considered. Thus our work may unveil the effects of incentives on cooperative governance of risky commons in a real scenario.

\hspace*{\fill}

\noindent Stimulating some nations or regions to reduce their greenhouse gas emissions is a fundamental task to mitigate climate change. How to design effective incentive measures is particularly important for this purpose. Since the Paris Agreement, we are witnessing the transformation of some countries' environmental governance to high-quality management~\citep{rogelj_nature16}. The establishment of the bonus-penalty mechanism, including giving appropriate compensation fees to emission reduction and imposing some fines on areas where pollution emission exceeds the prescribed quantity, has effectively promoted coordinated emission reduction~\citep{xue_jcp20}. However, it is still unclear how to weigh rewards and punishments in realistic yet complicated situations. Our research may provide some advices for policy makers from a theoretical perspective: pure reward is more effective than pure punishment and other hybrid incentive measures in achieving emission reduction targets, especially if the stability of the institution is of concern.

\section*{Limitations of the study}

\noindent The current study focuses on exploring the optimal incentive strategy in the scenario where the combination of punishment and reward is involved. However, it still has some limitations, which could be further extended. In this work, we consider the institution-type incentives with which punishment and reward are used. Indeed there are some other types of incentives in the real society and the implementation of incentive strategies is not necessarily institutional~\citep{vanlange_14,perc_pr17}. In the future work, we can thus consider different types of incentive strategies, e.g., social exclusion~\citep{Sasaki_prsb13} or different implementation ways of incentive, e.g., peer-type incentives~\citep{hauert_science07,sigmund_nature10}. In a recent work~\citep{han_plos21}, it is shown that peer punishment is more efficient than peer reward in promoting safety compliance and in mitigating the risk from Artificial Intelligence unsafe development. However, it is unknown whether pure reward strategy is still the best incentive strategy to promote the evolution of cooperation when peer-type incentives are considered in the collective-risk social dilemmas. In addition, in this work we consider that whether incentives can be provided or not depends on the number of executors. Indeed the number of cooperators can be used as a criteria to decide whether or not to provide incentives ~\citep{han_sr18}. In this scenario, it is interesting to explore the optimal incentive strategy for cooperative governance of risky commons. Moreover, implementation of local incentive requires additional information, and hence it is also necessary to consider the cost of information collection for executors. Therefore, it is worth further exploring how the introduction of observation cost or information collection cost affects cooperative governance of risky commons~\citep{szolnoki_pre15}. Our work reveals that pure reward strategy is a more efficient incentive strategy for cooperative governance of risky commons. This may result from that executors bear the cost of punishing defectors~\citep{szolnoki_prx17}, while the beneficiary of rewarding mechanism actually includes themselves, which needs to be verified in the future work. Indeed, considering second-order sanctions is important in the context of pool incentives~\citep{Fowler_nature05,perc_sr12,garcsif19}, so it may also open an interesting research avenue to explore the efficiency of incentives in the collective-risk social dilemma game. Previous studies have shown that the existence of second-order free-riders plays an important role in the emergence of cooperation and institutions~\citep{sigmund_nature10,sasaki_jtb11,szolnoki_prx17}. Therefore, it is worth investigating whether the introduction of second-order punishment and second-order reward can improve cooperative governance of risk commons~\citep{vanlange_14}. Furthermore, our research explores the optimal incentive strategy by adjusting the fixed relative weight parameter for the combination of punishment and reward. But indeed this parameter can be time-varying, depending on the population states. On the other hand, on the premise of ensuring a sufficiently good outcome, how to optimize the cost of providing incentives is another important issue of incentive design~\citep{han_sr18}. Hence it is meaningful to investigate the optimal incentive strategy in this scenario by means of optimal control theory or reinforcement learning approach~\citep{wang_cns19,ratliff_arcras19}. In this work, we have considered a well-mixed interaction where individuals perform rand interactions. However, the interactions among individuals are typically not random but rather that they are limited to a set of neighbors in a structured population, which could be described by a complex interaction network~\citep{santos_pnas11,szolnoki_pre11,pinheiro_njp12,szolnoki_prx13,perc_pr17}. Thus it could be interesting to explore the optimal incentive strategy in structured populations. We believe that future work considering these extensions will be valuable and improve our understanding of the role of incentive strategies in cooperative governance of risk commons.

\section*{Resource availability}

\section*{Lead contact}
\noindent Further information and requests for resources should be directed to and will be fulfilled by the Lead Contact, Xiaojie Chen (xiaojiechen@uestc.edu.cn).

\section*{Materials availability}
\noindent No materials were newly generated for this paper.

\section*{Data and code availability}
\noindent The data that support the results of this study are available from the corresponding authors upon request. The code used to generate the figures is available from the corresponding authors upon request.

\section*{STAR Methods}
\noindent Detailed methods are provided in the STAR Methods section of this paper and include Key Resources Table and Method Details.

\section*{ACKNOWLEDGMENTS}
\label{sec:4}
\noindent This work was supported by the National Natural Science
Foundation of China (Grant Nos. 61976048 and 62036002) and the Fundamental
Research Funds of the Central Universities of China.

\section*{AUTHOR CONTRIBUTIONS}
\noindent All authors designed and performed the research as well as wrote the paper.

\section*{DECLARATION OF INTERESTS}
\noindent The authors declare no competing interests.

\newpage

\noindent\textbf{STAR Methods}

\hspace*{\fill}

\noindent\textbf{Key Resources Table}

\begin{table}[H]
\label{tab1}
\begin{center}
\scriptsize
\begin{tabular}{cccccccccccc}\hline \textbf{REAGENT or RESOURCE} & \textbf{SOURCE} & \textbf{IDENTIFIER} \\\hline \textbf{Deposited data} \\\hline $r$ together with $b$, $c$, $Z$, $N$, $M$, $\beta$, and $\mu$ \\for collective-risk social dilemma in a finite population & Santos and Pacheco & https://doi.org/10.1073/pnas.1015648108 \\\hline \rowcolor{gray!25} $\pi_{t}$ together with $\pi_{e}$ and $n_E$ for fixed incentives & Vasconcelos {\it et al.} & https://doi.org/10.1038/nclimate1927 \\\hline $\delta$ together with $\pi_t$ for flexible incentives & Sasaki and Unemi & https://doi.org/10.1016/j.jtbi.2011.07.026 \\\hline
\textbf{Software and algorithms} \\\hline MATLAB R2014A & MathWorks & https://www.mathworks.com/ \\\hline
\end{tabular}
\end{center}
\end{table}

\hspace*{\fill}

\noindent\textbf{Method Details}

\hspace*{\fill}

\noindent \textbf{Collective-risk social dilemma game}

\noindent We consider a finite well-mixed population of \emph{Z} individuals. From the entire population, $N$ individuals are selected randomly to form a group for playing a collective-risk social dilemma game. In the group, each individual has an endowment $b$ at the beginning of game. Then, everyone decides whether or not to contribute an amount \emph{c} to the common pool. Furthermore, we require a minimum collective contribution $Mc$ ($0 < M \leq N$) or a minimum number $M$ of cooperators to ensure the benefit of everyone within the group. According to the traditional setup, cooperators contribute to reach the collective target while defectors not. If the collective target is not reached, all participants within the group will lose their remaining endowments with probability $r$ ($0 \leq r \leq 1$). Otherwise, individuals in the group retain their endowments. Notably, the parameter $r$ represents the risk level of collective failure \citep{santos_pnas11}.

\hspace*{\fill}

\noindent Based on the above description, the payoffs of defectors and cooperators in a group having $j_{C}$ cooperators and $N-j_{C}$ defectors obtained from the game can be respectively written as
\begin{eqnarray*}
\Pi'_{D}(j_{C}) & = & b\Theta(j_{C}-M)+(1-r)b[1-\Theta(j_{C}-M)]\,,\\
\Pi'_{C}(j_{C}) & = & \Pi'_{D}(j_{C})-c\,,
\end{eqnarray*}
where $\Theta(x)$ is the Heaviside function, that is, $\Theta(x)=0$ if $x < 0$, being one otherwise.

\hspace*{\fill}

\noindent In the following, we introduce a hybrid incentive strategy into the collective-risk social dilemma game and define two different kinds of enforcing institutions. Namely, in the first case the imposed incentives are fixed, while in the other case they are flexible.

\hspace*{\fill}

\noindent \textbf{Settings of fixed incentives}

\noindent We then introduce a third strategy into the collective-risk social dilemma game, that is, an executor (\emph{E}) who not only contributes an amount \emph{c} to the common pool, who but also contributes an investment to the sanctioning pool at a cost $\pi_{t}$. This makes it possible to punish free-riders and to reward cooperative players simultaneously. Accordingly, when the number of executors exceeds the given threshold $n_{E}$, each executor imposes an $\alpha$ portion of $\pi_{e}$ as positive incentive on every cooperators, while $1-\alpha$ portion of executive pool serves to punish free-riders. Here, $\alpha$ represents the relative weight between positive and negative incentives. In an extreme case, when $\alpha=0$, executors only choose to punish defectors and they become simple punishers. This simplified situation was discussed in a work of Vasconcelos {\it et al.}~\citep{vasconcelos_ncc13}. In the other extreme situation, when $\alpha = 1$, executors only choose to reward cooperators and \emph{E} becomes a simple rewarding strategy. To sum up our model, the payoffs of the competing strategies where $j_{C}$ cooperators and $j_{E}$ executors are present in the group can be written as
\begin{eqnarray*}
\Pi_{D}(j_{C},j_{E})&=&\Pi'_{D}(j_{C}+j_{E})-(1-\alpha)\pi_{e}\Delta,\\
\Pi_{C}(j_{C},j_{E})&=&\Pi'_{D}(j_{C}+j_E)+\alpha\pi_{e}\Delta-c,\\
\Pi_{E}(j_{C},j_{E})&=&\Pi'_{D}(j_{C}+j_E)+\alpha\pi_{e}\Delta-c-\pi_{t},
\end{eqnarray*}
where $\Delta$ corresponds to the sanction function, which relies on whether the sanction is local or global. When the local scheme is considered, $\Delta$ is given by $\Theta(j_{E}-n_{E})$, which means that the enforcement is executed at the group level if the number of executors in the group is not less than the group threshold. However, when the global scheme is considered, $\Delta = \Theta(i_{E}-n_{E})$, which means that incentives are applied in the whole population if the total number of executors $i_{E}$ is not less than the corresponding threshold level.

\hspace*{\fill}

\noindent \textbf{Settings of flexible incentives}

\noindent
Alternatively, the level of incentive $\pi_{e}$ may not be fixed but may depend on the number of executors. For instance, the institutional pool may behave as a typical public goods pool and its impact can be proportional to the simple sum of individual contributions $\pi_{e}$ of executor players with an enhancement factor $\delta > 1$. Notably, in the full reward extreme case this idea was used in one previous work~\citep{sasaki_jtb11}. More importantly, in our extended model we do not just apply a combined incentives strategy, but we still hold the collective-risk feature of the institutional pool. Therefore, as previously, the application of incentives depends sensitively on whether the $n_{E}$ threshold number of executors are present or not. Hence, the modified payoff values for flexible incentives are
\begin{eqnarray*}
\Pi_{D}(j_{C},j_{E})&=&\Pi'_{D}(j_{C}+j_{E})-(1-\alpha)\frac{\pi_{t}\delta j_{E}}{N-j_C-j_E}\Delta,\\
\Pi_{C}({j_{C},j_{E}})&=&\Pi'_{D}(j_{C}+j_{E})+\alpha\frac{\pi_{t}\delta j_{E}}{j_{C}+j_{E}}\Delta-c,\\
\Pi_{E}({j_{C},j_{E}})&=&\Pi'_{D}(j_{C}+j_{E})+\alpha\frac{\pi_{t}\delta j_{E}}{j_{C}+j_{E}}\Delta-c-\pi_{t}.
\end{eqnarray*}

\hspace*{\fill}

\noindent \textbf{Evolutionary dynamics}

\noindent
The average payoffs of the aforementioned strategies in a configuration $\textbf{i}=({i_{C},i_{E},i_{D}})$, characterized by the number of cooperators $i_C$, the number of executors $i_E$, and the number of defectors $i_D=Z-i_C-i_E$, can be computed by using a multivariate hypergeometric sampling \citep{hauert_science07} as follows
\begin{eqnarray*}
f_{D}&=&\sum\limits_{j_{C}=0}^{N-1}\sum\limits_{j_{E}=0}^{N-1-j_{C}}\frac{\binom{i_{C}}{j_{C}}\binom{i_{E}}{j_{E}}\binom{Z-i_{C}-i_{E}-1}{N-1-j_{C}-j_{E}}}{\binom{Z-1}{N-1}}\Pi_{D}(j_{C},j_{E}),\\
f_{C}&=&\sum\limits_{j_{C}=0}^{N-1}\sum\limits_{j_{E}=0}^{N-1-j_{C}}\frac{\binom{i_{C}-1}{j_{C}}\binom{i_{E}}{j_{E}}\binom{Z-i_{C}-i_{E}}{N-1-j_{C}-j_{E}}}{\binom{Z-1}{N-1}}\Pi_{C}(j_{C}+1,j_{E}),\\
f_{E}&=&\sum\limits_{j_{C}=0}^{N-1}\sum\limits_{j_{E}=0}^{N-1-j_{C}}\frac{\binom{i_{C}}{j_{C}}\binom{i_{E}-1 }{j_{E}}\binom{Z-i_{C}-i_{E}}{N-1-j_{C}-j_{E}}}{\binom{Z-1}{N-1}}\Pi_{E}(j_{C},j_{E}+1),
\end{eqnarray*}
where we note that $j_C\leq i_C$, $j_E\leq i_E$, and $j_D\leq i_D$ and thus impose that the binomial coefficients satisfy $\binom{i}{j}=0$ if $i<j$.

\hspace*{\fill}

\noindent To describe how individuals adopt new strategies over time, we consider a stochastic process of pairwise comparison~\citep{vansegbroeck_prl12}. In each evolutionary time step, a randomly chosen individual with strategy $L$ adopts the strategy $R$ of another randomly selected individual from the population with
probability $1/[1+e^{-\beta(f_{R}-f_{L})}]$, where the parameter $\beta$ corresponds to the intensity of selection~\citep{szabo_pre98}. For $\beta\rightarrow0$, we have random drift, while for $\beta\rightarrow \infty$ the imitation process is more inclined to prefer the strategies that yield higher payoffs~\citep{sigmund_nature10}. Without loss of generality, we set up $\beta=5.0$, meaning that it is very likely that the better performing players will pass their strategy to others, yet it is also possible that players will occasionally learn from a less successful individual. This setting is also consistent with the previous work~\citep{vasconcelos_ncc13}. We would like to stress that our main results remain valid when the value of $\beta$ is changed appropriately (see Figure S5 in Supplemental Information). In addition to the imitation, we also allow strategy mutation, and we assume that an individual adopts a different randomly selected strategy with probability $\mu$. Accordingly, the transition probability between strategies $R$ and $L$ under the above mutation-selection process can be written as
\begin{eqnarray}\label{transeq}
T_{L\rightarrow R}&=(1-\mu){\frac{i_{L}}{Z}\frac{i_{R}}{Z-1}[1+\mathop{{e}}\nolimits^{{-\beta(f_{R}-f_{L})}}]^{-1}}+\mu\frac{i_{L}}{2Z}\,,\nonumber
\end{eqnarray}
where $i_L$ ($i_R$) represents the number of individuals with strategy $L$ ($R$) in the population. Therefore, for a given configuration $\textbf{i}=({i_{C},i_{E},i_{D}})$, the probability that the number of cooperators increases or decreases by one in the entire population is written as
\begin{equation*}
T_{\textbf{i}}^{C\pm}=T_{(i_{C}\pm1,i_{E}\mp1,i_{D})}+T_{(i_{C}\pm1,i_{E},i_{D}\mp1)}\,.
\end{equation*}
Similarly, the probability that the number of executors increases or decreases by one is given by
\begin{equation*}
T_{\textbf{i}}^{E\pm}=T_{(i_{C}\mp1,i_{E}\pm1,i_{D})}+T_{(i_{C},i_{E}\pm1,i_{D}\mp1)}\,.
\end{equation*}

\hspace*{\fill}

\noindent Since the strategy update process depends only on the current configuration of the system, the associated evolutionary dynamics can be described as a Markov process over a two-dimensional space~\citep{kampen07} where the probability density function we designate by $p_{\textbf{i}}(t)$, satisfies the following Master Equation
\begin{flalign*}
&&p_{\textbf{i}}(t+\tau)-p_{\textbf{i}}(t)&=\sum_{\textbf{i}^{'}}\left\{T_{\textbf{i}\textbf{i}^{'}}p_{\textbf{i}^{'}}(t)-T_{\textbf{i}^{'}\textbf{i}}p_{\textbf{i}}(t)\right\}\,.&
\end{flalign*}
Here $T_{\textbf{i}\textbf{i}^{'}}$ denotes the transition probability from the configuration $\textbf{i}'$ to $\textbf{i}$ per unit time $\tau$~\citep{imhof_pnas05}. Accordingly, we can obtain the so-called stationary distribution $\bar{p}_{\textbf{i}}(t)$, by searching the eigenvector associated with the eigenvalue 1 of the transition matrix $\textbf{T}=[T_{\textbf{ij}}]^{T}$ \citep{kampen07}. As an important quantity for describing the evolutionary dynamics of strategies in the system, the stationary distribution can characterize the fraction of time that the population spends in each possible configuration of the finite population.

\hspace*{\fill}

\noindent In addition to the stationary distribution $\bar{p}_{\textbf{i}}(t)$, another
central quantity for describing the evolutionary dynamics of strategies is the gradient of selection $\nabla_{\textbf{i}}$, which describes the most likely evolutionary path of change of population configuration with time. Employing the Kramers-Moyal expansion of the Master Equation \citep{helbing_pa93}, we can obtain the gradient of selection as
\begin{flalign*}
&&\nabla_{\textbf{i}}&=(T_{\textbf{i}}^{C+}-T_{\textbf{i}}^{C-})\textbf{u}_{\textbf{C}}+(T_{\textbf{i}}^{E+}-T_{\textbf{i}}^{E-})\textbf{u}_{\textbf{E}}\,,&
\end{flalign*}
where $\textbf{u}_{\textbf{C}}$ and $\textbf{u}_{\textbf{E}}$ are unit vectors defining as a basis of the two dimensional simplex.

\hspace*{\fill}

\noindent Furthermore, in order to investigate the role of different sanctioning policies, we provide a key quantity $a_{G}(\textbf{i})$ to compute the fraction of groups that reach a given threshold, that is, the collective target~\citep{vasconcelos_ncc13}. We should compute the average group achievement from the perspective of the entire population and not obtain the value for groups centered on a given $C$, $E$, or $D$.
For each possible configuration $\textbf{i}=(i_{C},i_{E},i_{D})$, accordingly we adopt the multivariate hypergeometric distribution to calculate $a_{G}(\textbf{i})$ as
\begin{flalign*}
&&a_{G}(\textbf{i})&\!=\!\tbinom{Z}{N}^{-1}\sum\limits_{j_{C}\!=\!0}^{N}\sum\limits_{j_{E}\!=\!0}^{N-j_{C}}\tbinom{i_{C}}{j_{C}}\tbinom{i_{E}}{j_{E}}\tbinom{Z-i_{C}-i_{E}}{N-j_{C}-j_{E}}\Theta(j_{C}\!+\!j_{E}\!-\!M)\,.&
\end{flalign*}
Subsequently, by using $a_{G}(\textbf{i})$ combined with the stationary distribution $\bar{p}_{\textbf{i}}(t)$, we can calculate the average group achievement $\eta_{G}$ as an important quantity which refers to the average proportion of groups that achieve the collective target in all possible group configurations, that is, $\eta_{G}=\sum_{\textbf{i}}\bar{p}_{\textbf{i}}a_{G}(\textbf{i})$.

\hspace*{\fill}

\noindent We further compute the fraction of groups that reach $n_E$ executors for local institution or whether a global institution can be formed for the configuration \textbf{i}. Accordingly, we should compute the institution prevalence from the perspective of the entire population and not obtain the value for groups centered on a given $C$, $E$, and $D$. In both cases, we designate this quantity by $a_{I}(\textbf{i})$, which can be calculated as
\begin{flalign*}
&&a_{I}(\textbf{i})&\!=\!\tbinom{Z}{N}^{-1}\sum\limits_{j_{C}\!=\!0}^{N}\sum\limits_{j_{E}\!=\!0}^{N-j_{C}}\tbinom{i_{C}}{j_{C}}\tbinom{i_{E}}{j_{E}}\tbinom{Z-i_{C}-i_{E}}{N-j_{C}-j_{E}}\Delta\,.&
\end{flalign*}
We subsequently use $a_{I}(\textbf{i})$ combined with the stationary distribution $\bar{p}_{\textbf{i}}(t)$ to compute the institution prevalence $\eta_{I}$ averaging over all possible configuration \textbf{i}, which is given as $\eta_{I}=\sum_{\textbf{i}}\bar{p}_{\textbf{i}}a_{I}(\textbf{i})$.

\renewcommand{\figurename}{\textbf{Figure~S}}

\newpage

\centerline{Supplemental Information for}
\vbox{}

\centerline{\normalsize\textbf{Combination of institutional incentives for cooperative governance of risky commons}}
\vbox{}

\centerline{Weiwei Sun, Linjie Liu, Xiaojie Chen, Attila Szolnoki, and V\'{\i}tor V. Vasconcelos}

\vbox{}

Here we present detailed numerical results about the evolutionary dynamics in finite populations when fixed and flexible incentives are respectively considered under global scheme of sanctioning.

\section*{Fixed incentives under global scheme}

\setcounter{figure}{0}
\begin{figure*}[b!]
\begin{center}
\includegraphics[width=12cm]{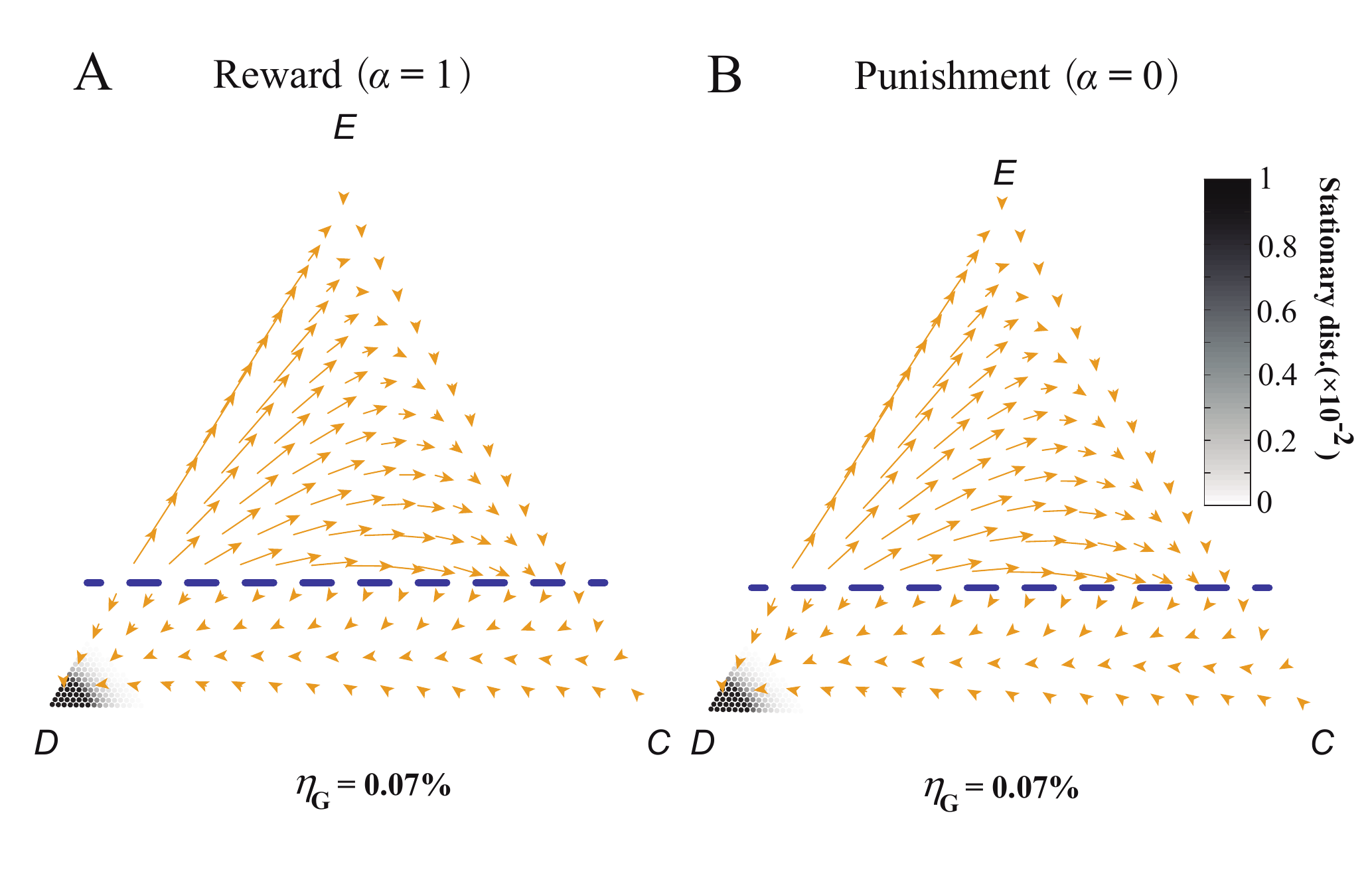}
\caption{\textbf{Evolutionary dynamics of cooperators, defectors, and executors under global scheme with fixed incentives.} The risk level is \emph{r} = 0.2. In panel A, the $\alpha = 1$ extreme case of pure reward strategy is considered. In panel B, the other $\alpha = 0$ extreme case of pure punishment is considered. Incentives are imposed on global level when the number of executors exceeds $25\%$ of the population. Parameter values are \emph{Z} = 100, \emph{N} = 4, \emph{c} = 0.1, \emph{b} = 1, $\mu=1/Z$, $\pi_{t}=0.03$, and $\pi_{e}=0.3$.
}\label{figS1}
\end{center}
\end{figure*}

\noindent In Figure~S\ref{figS1} we first show the evolutionary dynamics of cooperators, defectors, and executors in the entire population under global scheme when fixed values of incentives are used. We find that when executors choose to reward cooperators ($\alpha=1$) or punish defectors ($\alpha=0$) exclusively, then the entire population always spends most time near the full defection state. The direction flow of orange arrows suggests that when the number of executors is larger than a given threshold (see blue dash line), most arrows flow to the \emph{CE}-edge, and then enter the bottom of the blue dashed line along the \emph{CE}-edge, and finally arrive to the vertex \emph{D}. Accordingly, free-riders can dominate the entire population no matter whether pure reward strategy or pure punishment strategy is considered under global scheme. Furthermore, in order to better distinguish the effectiveness of pure reward strategy and pure punishment strategy, we have calculated the group achievement values for $\alpha=1$ and $\alpha=0$, and we have obtained that both values are $0.07\%$. Hence, we find that pure reward strategy does not have a distinct advantage over pure punishment strategy in improving the level of group achievement. Indeed this is because the payoff differences between any two competing strategies are independent of the parameter $\alpha$ for fixed incentives under global scheme, and then we can have the same stationary distribution and the same gradient of selection for different values of $\alpha$.

\begin{figure*}[b!]
\begin{center}
\includegraphics[width=12cm]{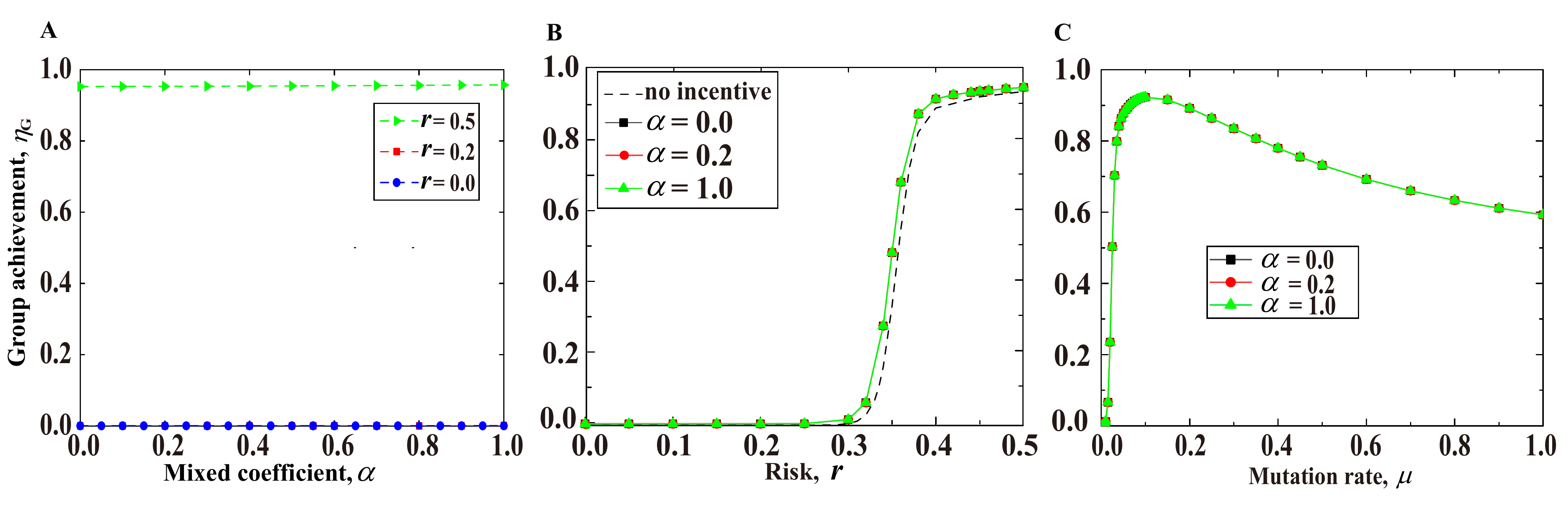}
\caption{\textbf{The average group achievement $\eta_{G}$ under global scheme with fixed incentives.} Parameter values are \emph{Z} = 100, \emph{N} = 4, \emph{c} = 0.1, \emph{b} = 1, $\mu=1/Z$, $\pi_{t}=0.03$, and $\pi_{e}=0.3$ in panels~A and B; \emph{Z} = 100, \emph{N} = 4, \emph{c} = 0.1, \emph{b} = 1, \emph{r} = 0.3, $\pi_{t}=0.03$, and $\pi_{e}=0.3$ in panel~C.}\label{figS2}
\end{center}
\end{figure*}

\hspace*{\fill}

\noindent Figure~S\ref{figS2} summarizes how the group achievement $\eta_{G}$ depends on the key parameters of the model. In panel~A we show that there is no distinct change in dependence of $\alpha$. Hence, this suggests that pure reward strategy does not improve the level of group achievement than other combination of incentives for different values of $r$. Panel~B of Figure~S\ref{figS2} demonstrates clearly that higher risk level enforces higher group achievement. For comparison, we here also show the two-strategy case when incentive strategy is missing (see black dash line). We can find that the introduction of $E$ strategy improves the results. Last we present the effect of mutation rate $\mu$ on the group achievement $\eta_{G}$ in panel~C of Figure~S\ref{figS2}. Similarly to local institutions, the value of $\eta_{G}$ first increases, then reaches a maximum, and finally saturates for high $\mu$ values. However, we still find that for fixed incentives under global scheme pure reward strategy does not have an evident advantage over other incentive strategies when strategy $E$ rewards cooperators exclusively.

\section*{Flexible incentives under global scheme}

\begin{figure*}[t]
\begin{center}
\includegraphics[width=11cm]{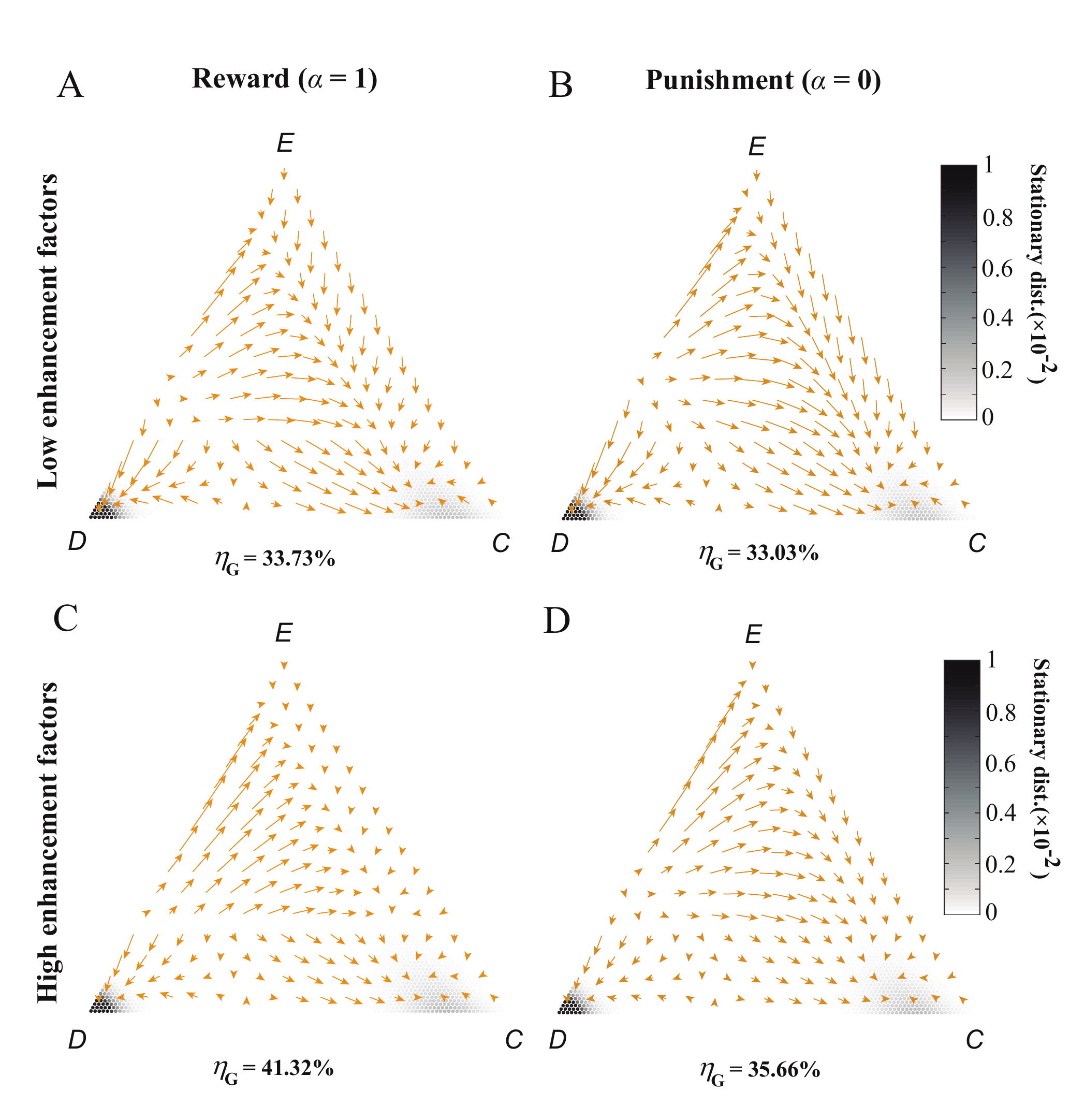}
\caption{\textbf{Evolutionary dynamics of cooperator, defector, and executor strategy under global scheme with flexible incentives.} The risk level is \emph{r} = 0.35. Panels~A and C show the case of pure reward ($\alpha=1$). Panels~B and D show the case of pure punishment ($\alpha=0$). Parameter values are \emph{Z} = 100, \emph{N} = 4, \emph{c} = 0.1, \emph{b} = 1, $\mu=1/Z$, $\pi_{t}=0.03$, and $\delta=1.4$ in panels~A and B; \emph{Z} = 100, \emph{N} = 4, \emph{c} = 0.1, \emph{b} = 1, $\mu = 1/Z$, $\pi_{t}=0.03$, and $\delta=3$ in panels~C and D.}\label{figS3}
\end{center}
\end{figure*}

\noindent We first show the effectiveness of pure reward strategy and pure punishment strategy in improving the level of group achievement, as shown in Figure~S\ref{figS3}. Importantly, the imposed incentives are not fixed, instead the incentive values depend on the number of executors, whose collective efforts are acknowledged via an enhancement factor $\delta$. The comparison of simplex suggests that there is no qualitative difference between the cases of low and high $\delta$ values. Namely, the population not only spends most of the time in the vicinity of vertex \emph{C}, but also spends a reasonable time very close to vertex \emph{D}. This is valid both for pure reward and pure punishment strategies. For a more accurate comparison, we have calculated the group achievement values for both cases and added their values to the simplexes. Accordingly, this value is always higher for $\alpha=1$, regardless of the enhancement factor values.

\hspace*{\fill}

\begin{figure*}[t]
\begin{center}
\includegraphics[width=11cm]{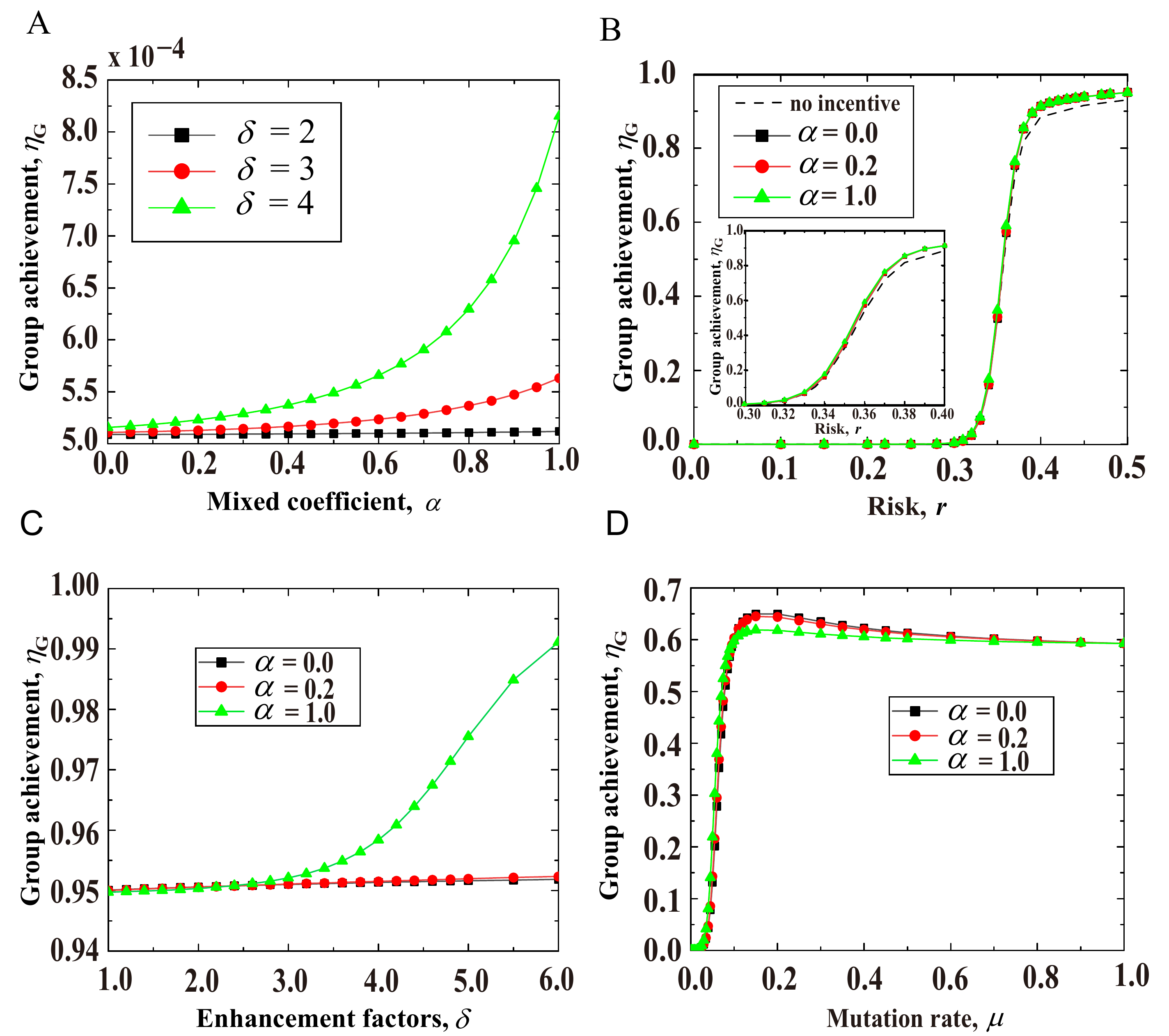}
\caption{\textbf{The average group achievement $\eta_{G}$ under global scheme with flexible incentives.} The applied weight factor values are marked in the legends and panel~B also contains the traditional case in the absence of strategy $E$. Parameter values are \emph{Z} = 100, \emph{N} = 4, \emph{c} = 0.1, \emph{b} = 1, $\mu = 1/Z$, \emph{r} = 0.2, and $\pi_{t} = 0.03$ in panel~A; \emph{Z} = 100, \emph{N} = 4, \emph{c} = 0.1, \emph{b} = 1, $\mu = 1/Z$, $\delta = 2$, and $\pi_{t} = 0.03$ in panel~B; \emph{Z} = 100, \emph{N} = 4, \emph{c} = 0.1, \emph{b} = 1, $\mu = 1/Z$, $\emph{r} = 0.5$, and $\pi_{t} = 0.03$ in panel~C; \emph{Z} = 100, \emph{N} = 4, \emph{c} = 0.1, \emph{b} = 1, $\mu = 1/Z$, \emph{r} = 0.2, $\delta = 3$, and $\pi_{t} = 0.03$ in panel~D.}\label{figS4}
\end{center}
\end{figure*}

\noindent To explore the proper consequence of mixed incentives, we present $\eta_{G}$ as a function of $\alpha$ for three different values of $\delta$ in Figure~S\ref{figS4}A. Similarly to the local scheme case, we find that when the value of $\delta$ is small, the level of group achievement is very low, regardless of the value of $\alpha$. By increasing $\delta$, however, $\eta_{G}$ can be improved significantly for large $\alpha$. Hence we can conclude that pure reward strategy is always more effective than other combinations of incentives, especially when the value of $\delta$ is large.
In Figure~S\ref{figS4}B we present $\eta_{G}$ as a function of $r$ for three different values of $\alpha$. For comparison, we also show the group achievement in the absence of executor strategy. As expected, $\eta_{G}$ grows by increasing the risk level of collective failure, but the curves evolve in a very similar way. Strictly speaking, the curve belonging to $\alpha=1$ case is always higher than other curves, hence the superiority of pure reward option remains intact. Next we present $\eta_{G}$ as a function of $\delta$ for three different values of $\alpha$. Figure~S\ref{figS4}C suggests that the group achievement level increases by increasing the value of $\delta$. Furthermore, the usage of pure reward strategy can be more effective than any other mixture of incentives especially when the value of $\delta$ is large. Finally we show how the mutation rate $\mu$ influences the group achievement $\eta_{G}$ for three different values of $\alpha$. Figure~S\ref{figS4}D shows that the advantage of pure reward against any other combinations is justified when the mutation is less than $0.2$ for flexible incentives under global scheme.

\section*{Robustness to the intensity of selection }

\begin{figure*}[t]
\begin{center}
\includegraphics[width=12cm]{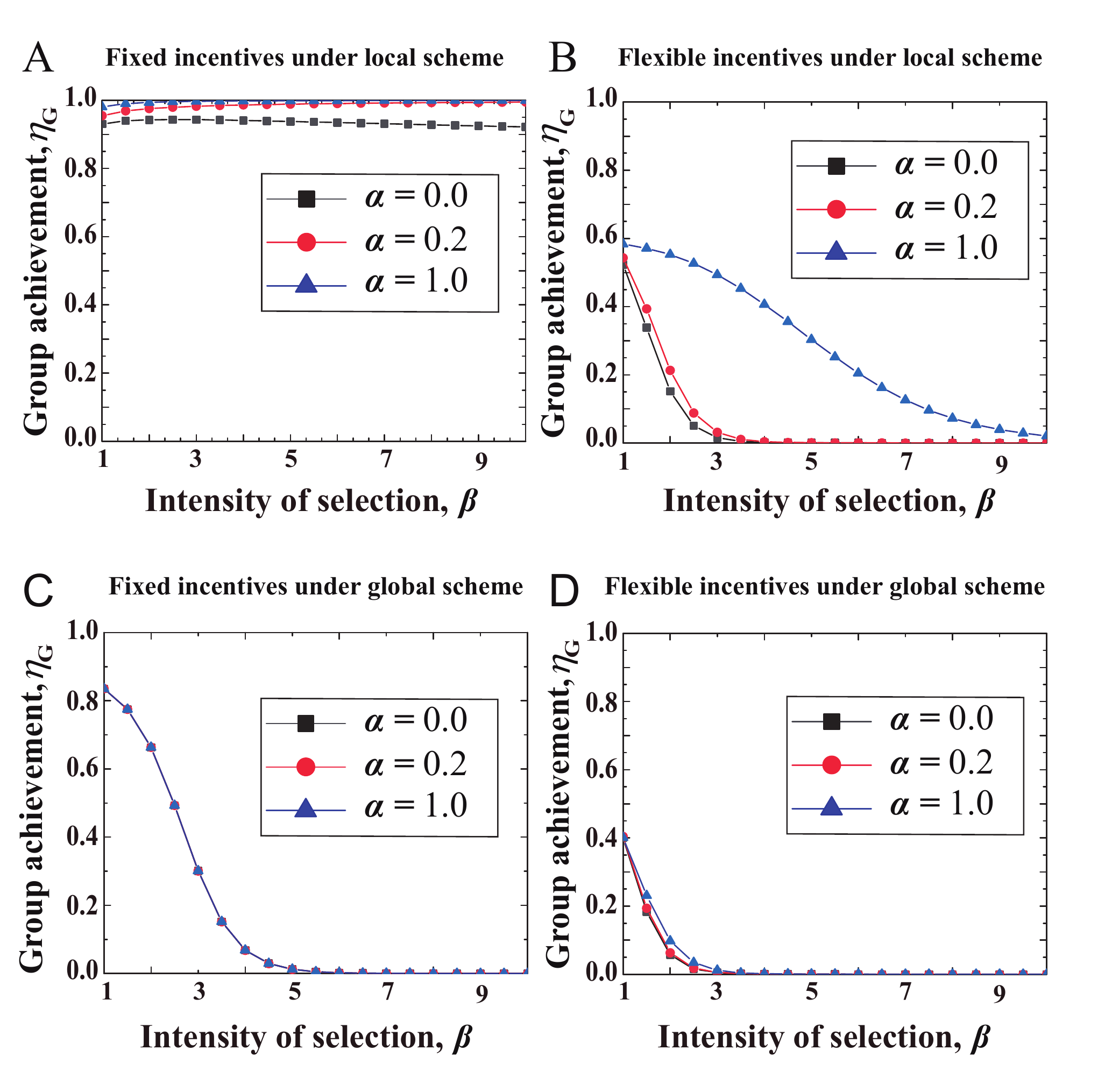}
\caption{\textbf{Robustness of the average group achievement $\eta_{G}$ to the changes of the intensity of selection.} Panel A shows $\eta_{G}$ as a function of $\beta$ for three different values of $\alpha$ under local scheme with fixed incentives. Panel B shows $\eta_{G}$ as a function of $\beta$ for three different values of $\alpha$ under local scheme with flexible incentives. Panel C shows $\eta_{G}$ as a function of $\beta$ for three different values of $\alpha$ under global scheme with fixed incentives. Panel D shows $\eta_{G}$ as a function of $\beta$ for three different values of $\alpha$ under global scheme with flexible incentives. Parameter values are $Z=100$, $N=4$, $c=0.1$, $b=1$, $r=0.3$, $\pi_{t}=0.03$, and $\pi_{e}=0.3$ in panel~A; $Z=100$, $N=4$, $c=0.1$, $b=1$, $r=0.2$, $\pi_{t}=0.03$, and $\delta=3$ in panel~B; $Z=100$, $N=4$, $c=0.1$, $b=1$, $r=0.3$, $\pi_{t}=0.03$, and $\pi_{e}=0.3$ in panel~C; $Z=100$, $N=4$, $c=0.1$, $b=1$, $r=0.2$, $\pi_{t}=0.03$, and $\delta=3$ in panel~D.}\label{figS5}
\end{center}
\end{figure*}

In Figure~S\ref{figS5}, we show how the group achievement varies when the intensity of selection is changed appropriately for fixed incentives under local scheme, flexible incentives under local scheme, fixed incentives under global scheme, and flexible incentives under global scheme, respectively. We find that our main results remain valid when changing the value of intensity of selection, that is, pure reward strategy outperforms other combinations of incentives for fixed incentives under local scheme, flexible incentives under local scheme, and flexible incentives under global scheme. However, for fixed incentives under global scheme, we find that changing the value of intensity of selection can alter the group achievement for each $\alpha$ value, but there is no distinct change in dependence of $\alpha$ when the value of intensity of selection is fixed.

\end{document}